\documentclass[aps,reprint,groupedaddress,superscriptaddress,showpacs]{revtex4-2}

\usepackage{graphicx}
\usepackage{dcolumn}
\usepackage{bm}
\usepackage{hyperref}
\usepackage{amsmath,amsfonts}
\usepackage{textcomp,gensymb}
\renewcommand*{\eqref}[1]{\hyperref[{#1}]{\textup{(\ref*{#1})}}}
\newcommand*{\figref}[1]{\hyperref[{#1}]{\textup{Fig.~\ref*{#1}}}}
\newcommand*{\mat}[1]{\overline{\overline#1}}

\begin{document}

\title{Going Beyond Perfect Absorption: Reconfigurable Super-directive Absorbers}

\author{Yongming Li}
\email[]{yongmingli@stu.xjtu.edu.cn}
\affiliation{State Key Laboratory of Electrical Insulation and Power Equipment, School of Electrical Engineering, 
Xi'an Jiaotong University, Xi'an 710049, China}
\affiliation{Department of Electronics
and Nanoengineering, Aalto University, P.O. Box 15500, FI-00076, Espoo, Finland}
\author{Xikui Ma}
\affiliation{State Key Laboratory of Electrical Insulation and Power Equipment, School of Electrical Engineering, 
Xi'an Jiaotong University, Xi'an 710049, China}
\author{Xuchen Wang}
\affiliation{Institute of Nanotechnology, Karlsruhe Institute of Technology, P.O. Box 3640, 76021, Karlsruhe, Germany}
\author{Sergei A. Tretyakov}
\affiliation{Department of Electronics
and Nanoengineering, Aalto University, P.O. Box 15500, FI-00076, Espoo, Finland}

\date{\today}

\begin{abstract}
In the context of electromagnetic absorption, it is obvious that for an infinite planar periodic structure illuminated by a plane wave, the maximum attainable absorptance, i.e., perfect absorption, is theoretically limited to 100\% of the incident power. Here we show that an intriguing possibility of overcoming this limit arises in finite-size resonant absorbing arrays.  We present a comprehensive analysis of a simple two-dimensional strip array over an infinite perfectly conducting plane, where the strips are loaded by reconfigurable impedance loads. The absorptance is defined as the ratio of the dissipated power per unit length of the strips to the incident power on the unit length of the array width.  The results show that even regular arrays of impedance strips can slightly overcome the limit of 100\% absorptance, while using aperiodic arrays with optimized loads, absorptance can be significantly increased as compared with the scenario where the strips are identical. In principle, by tuning the reconfigurable loads, high super-unity absorptance can be realized for \textit{all} angles of illumination. 
\end{abstract}


\maketitle

\section{Introduction}
Absorbers for electromagnetic waves have a pivotal role in various applications, such as, for example, energy harvesting~\cite{Xie2017wideband}, stealth technology~\cite{Wen2022ultrabroadband,Fu2023rcs}, and sensors~\cite{Rakhshani2022narrowband,Singh2022designing}. In the literature, absorbers are called \emph{perfect} if they absorb all the power incident on their surface, at least at a specific frequency and the angle of incidence. Motivated by the variety of applications, a lot of studies on perfect absorbers have been conducted, in particular, exploiting metasurfaces or metamaterials. Ref.~\cite{Ra2015thin} provides a tutorial overview on the phenomenon of perfect absorption in infinite optically thin planar layers and classifies perfect absorbers according to their operational principles. For an infinite periodic structure illuminated by a propagating plane wave, the maximum absorptance is obviously $100\%$.
However, the absorptance of finite-sized bodies can sometimes exceed $100\%$. This typically occurs when these absorbers work at the conjugate impedance match condition to maximize the received power~\cite{Maslovski2016Overcoming,valagiannopoulos2015electromagnetic}. Higher than 100\% absorptance means that the absorber can capture and absorb more power than is incident on its geometric cross section. This means that the ideal black body~\cite{Kirchhoff1860relation} that was introduced by Kirchhoff in 1860, is not the ultimate absorber, although it absorbs all the incident rays falling on its surface. In particular, it is well known that small resonant particles, for example, subwavelength-sized metallic particles, are able to absorb significantly more than a black body of the same size, e.g.~\cite{tretyakov2014maximizing,Maslovski2016Overcoming, Maslovski2018Superabsorbing}.  Essentially, these resonant particles have the capability to gather the power of the incident wave from an area significantly exceeding the physical cross section of the particles. Similarly, in the antenna theory, it is known that the upper limit of the effective area of a resonant dipole is ${3\over8\pi}\lambda^2$, where $\lambda$ is the corresponding working  wavelength~\cite{tretyakov2014maximizing}. This value is independent from the dipole size, which means that no upper limit of the absorption cross section exists if multipolar resonant modes of the object are permitted, e.g.~\cite{Fan,Mann,Mohammadi2014Minimum,Maslovski2018Superabsorbing}. The majority of superabsorption research focuses on small particles rather than
on electrically large bodies. While it is known that theoretically there is no limit on how large the absorption cross section can be, and some approaches to physical realizations of large super-absorbing bodies have been proposed, e.g. \cite{Maslovski2018Superabsorbing,valagiannopoulos2015electromagnetic}, that would require filling bulk bodies by complex media with highly-resonant
and extremely low-loss microstructures. 

Conventional realizations of thin resonant absorbers are based on metasurfaces, most commonly in the form of 
multi-element resonant arrays \cite{Ra2015thin}. It is believed that such electrically large but finite in size absorbing metasurfaces and multi-element resonant arrays perform similarly to corresponding infinite-sized structures in terms of absorption efficiency, absorbing a maximum of 100\% of the power incident on their surfaces. Truncated (finite-size) periodic structures, as a specific type of multi-element resonant arrays, are often regarded as practically the simplest and most effective realizations of thin resonant absorbers.  Researchers have rarely explored the difference between infinite-sized periodic absorbers and finite-size, truncated ones. Moreover, to the best of our knowledge, it is not known if it is possible to increase the effective area above 100\% of its geometrical size and what is required for the realization of such extreme properties of absorbing metasurfaces.   

In this work, we examine a simple, analytically solvable example of a  two-dimensional array of impedance strips and present two distinctive absorbers based on finite-length arrays. One scenario concerns the truncated periodic array where all the strips are loaded by identical impedance loads, while the other focuses on arrays with globally optimized load impedances. Both of these  absorbers exhibit absorptance that exceeds 100\% for a certain angle of plane-wave incidence. Our results show that optimization of load impedances globally can enable absorption of more incident power as compared to conventional finite-width regular arrays. This research reveals that it is possible for absorber designs to go beyond ``perfect absorption" in electrically thin metasurfaces.

\section{Principle and methodology}

We consider geometrically periodic strip arrays placed over a perfectly conducting (PEC) ground plane at a distance $h$. The strips are periodically loaded by impedance loads with the impedances per unit length equal to $Z_{\rm L}$ [$\Omega\text{/m}$]. The distance between the insertions  $l$ is electrically small, so that the loaded strips can be considered as effectively uniform impedance strips. 

\begin{figure}[!htbp]
    \centering
    \includegraphics{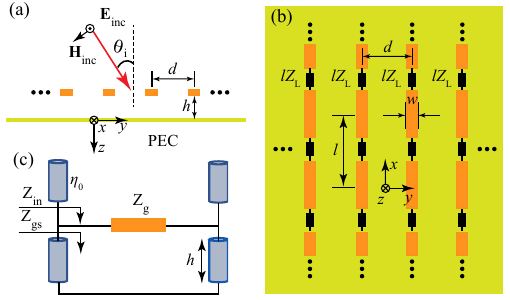}
    \caption{(a) Front view of an infinite periodic strip array placed over an infinite ground plane and illuminated by a plane wave traveling in the direction of $\theta_{\rm i}$. The distance between the two adjacent strips is $d$. (b) Top view of the array. The strips are loaded by bulk impedances inserted periodically with the period $l$. The width of the strips is $w$.  Both $l$ and $w$ are much smaller than the wavelength in free space. The periodically loaded strips can be modeled as homogeneous impedance strips with the impedances per unit length  $Z_{{\rm L}}$.  (c) Equivalent circuit of the system.}
    \label{fig:inf_configuration}
\end{figure}

As the reference case and the initial design step, we first consider an infinite periodic array, where all the load impedances $Z_{\rm L}$ are identical. The array is illuminated by a TE polarized plane wave $\mathbf{E}_{\rm inc} = E_0 e^{-j k_0 \sin \theta_{\rm i} y -j k_0 \cos \theta_{\rm i} z} \hat{x}$, where $k_0=\omega_0 \sqrt{\epsilon_0 \mu_0}$ is the wave number in free space, see an illustration in Figs.~\ref{fig:inf_configuration}(a) and (b). The reflected wave from the ground plane is given by $\mathbf{E}_{\rm ref} = -E_0 e^{-j k_0 \sin \theta_{\rm i} y + j k_0 \cos \theta_{\rm i} z} \hat{x}$. 
This is one of the simplest examples of thin metasurface absorbers, see \cite{Ra2015thin}. 
For such an infinite-sized periodic structure, transmission-line theory can be conveniently used~\cite{tretyakov2003analytical}, and the equivalent circuit is depicted in~\figref{fig:inf_configuration}(c).

For simplicity of analysis, we assume that the array is in free space. According to Ref.~\cite{pozar2011microwave}, the input impedance of the grounded substrate is given by $Z_{\rm gs} = j \tilde{Z}_0 \tan \left( k_0 \cos \theta_{\rm i} h \right)$, where $\tilde{Z}_0 = \frac{\eta_0}{\cos \theta_{\rm i}}$ represents the characteristic impedance for TE-polarized plane waves for an incident angle of $\theta_{\rm i}$. $\eta_0$ is the free-space impedance. The input impedance of the whole structure is the parallel connection of $Z_{\rm gs}$ and the grid impedance $Z_{\rm g}$ of the strip array, that is, $Z_{\rm in} = Z_{\rm gs} \parallel Z_{\rm g}$. The equivalent grid impedance of dense periodic strips array reads~\cite[Eq.~4.38]{tretyakov2003analytical}
\begin{equation}
    Z_{\rm g} = Z_{\rm L} d + j \frac{\eta_0}{2} \alpha_{\rm ABC},
\end{equation}
where the grid parameter $\alpha_{\rm ABC} = \frac{k_0 d}{\pi} \log \frac{d}{2 \pi r_{\rm eff}}$ (see~\cite[Eq.~4.32]{tretyakov2003analytical}), and the effective radius is given by $r_{\rm eff}=\frac{w}{4}$. To realize perfect absorption, the input impedance of the infinite structure should match the impedance of free space for the incident wave. The required load impedance for the design incident angle $\theta_{\rm i}$ is determined by
\begin{equation}
    Z_{\rm L} = \frac{1}{d} \left( \frac{Z_{\rm gs}\tilde{Z}_0 }{Z_{\rm gs} - \tilde{Z}_0} - j \frac{\eta_0}{2} \alpha_{\rm ABC} \right).    \label{eq:periodic_load_imp_den}
\end{equation}

For an infinite strip array, the induced currents flowing on the strips are calculated by  
\begin{equation}
    I_{\rm inf} = \frac{2 \left( E_{\rm ref} - R E_{\rm inc} \right)}{\tilde{Z}_0 \left( 1-e^{-j k_0 \cos \theta_{\rm i} 2 h}\right)} d,
    \label{eq:inf_induced_current}
\end{equation}
where the reflection coefficient $R$ for TE polarized plane wave reads $\smallskip{R= \left( Z_{\rm in} - \tilde{Z}_0 \right)  \left( Z_{\rm in} + \tilde{Z}_0 \right)^{-1} }$. The induced currents at infinite strip array absorbers, given by \eqref{eq:inf_induced_current}, will be later utilized for comparison with finite-sized absorbers.

\begin{figure}[!htbp]
    \centering    \includegraphics{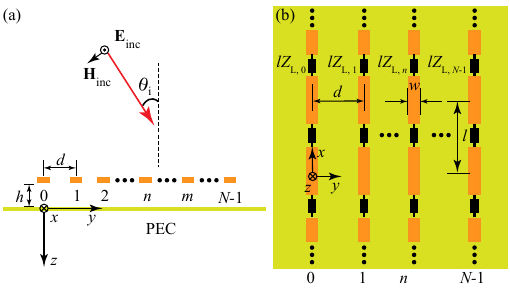}
    \caption{(a) A finite-width strip array above an infinite PEC ground plane  under illumination by a TE-polarized plane wave at $\theta_{\rm i}$. (b) Top view of the structure. The first strip is at the position  $y=0$, $z=-h$.}    \label{fig:finite_sized_config}
\end{figure}

In practice, designed  periodic arrays need to be truncated to finite-size structures. 
The corresponding finite-size strip array is shown in Figs.~\ref{fig:finite_sized_config}(a) and (b). In finite arrays, it makes sense to use different load impedances for different strips as a mean to optimize absorption. The truncated periodic array is a special case when all the loads are still the same,  $Z_{{\rm L},n} = Z_{\rm L}$ with $\{n = {0,1, \dots, N-1}\}$. When such a finite-sized structure is excited by a plane wave, the induced current $I_m$ that flows on the surface of the $m$-th strip dissipates Joule heat in lossy loads. The absorptance $A$ of the finite-sized structure we define as  
\begin{equation}
    A = \frac{P_{\rm dis}}{P_{\rm inc}},
    \label{eq:absorption}
\end{equation}
where the dissipated power is 
\begin{equation}
    P_{\rm dis} = \frac{1}{2}\sum_{m=0}^{N-1} |I_m|^2 \Re\{ Z_{{\rm L},m} \}
    \label{eq:dissipated_power}
\end{equation} 
and the incident power on the array's geometric area (per unit length along  $x$) is $P_{\rm inc} = {E_0^2 N d \over 2 \tilde{Z}_0}$. This definition is used to evaluate whether the designed absorber has a super-directive property or not. If the absorptance is larger than 100\%, it means that the designed absorber is a super-directive absorber, as it absorbs more power than is falling on its surface. 

For finite-sized absorbing arrays, the total external electric field $E_x^{\rm ext} (y,z) =  -j 2 E_0 \sin (k_0 \cos \theta_{\rm i} z) e^{- j k_0 \sin\theta_{\rm i} y}$ at the coordinate $(y,z)$ is the superposition of the incident wave and its reflection from the ground plane. For a given set of load impedances $\vec{Z}_{\rm L}=(Z_{{\rm L}, 0}, Z_{{\rm L}, 1},\dots, Z_{{\rm L}, N-1})^{\rm T}$, the induced current can be easily obtained by a simple matrix operation according to~\cite[Eq.~(5)]{li2023all}
\begin{equation}
     \vec{I} = \mat{Z}^{-1} \cdot \vec{U},
    \label{eq:finite_induced_current}
\end{equation}
where the column vector of the induced currents is represented by $\smallskip{ \vec{I} = \left[ I_0, I_1, \dots, I_{N-1} \right]^{\rm T} }$, while the total external voltage vector is represented by $\smallskip{ \vec{U} =  \left[ E^{\rm ext}_x(y_0,-h), E^{\rm ext}_x(y_1,-h), \dots, E^{\rm ext}_x(y_{N-1},-h) \right]^{\rm T} }$. $\mat{Z} = \mat{Z}_{\rm s} + \mat{Z}_{\rm m} +\mat{Z}_{\rm L}$ is the impedance matrix that is composed of the self-impedance matrix (a diagonal matrix) $\mat{Z}_{\rm s}={\rm diag}\left(Z_0, Z_1, \dots, Z_n, \dots, Z_{N-1} \right)$, the load impedance matrix $\mat{Z}_{\rm L} = {\rm diag}(\vec{Z}_{\rm L})$, and the mutual impedance matrix $\mat{Z}_{\rm m}$. The entry of the self impedance matrix $Z_n={k_0 \eta \over 4} \left[H_0^{(2)}(k_0 r_{\rm eff}) - H_0^{(2)}(2 k_0 h) \right]$ is the self-impedance of strips $n$, and the entry of the mutual impedance $Z_{nm} = \frac{k_0 \eta}{4} \left [ H_0^{(2)} \left(k_0 \left| y_m-y_n \right| \right) - H_0^{(2)} \left( k_0 \sqrt{(y_m-y_n)^2 + 4h^2} \right) \right] $ is the mutual impedance between strips $m$ and $n$. After knowing the induced currents, the absorptance and dissipated power can be found from~ \eqref{eq:absorption} and \eqref{eq:dissipated_power}, respectively. In this work, the example operation frequency is chosen as $f_0=10~\text{GHz}$. The distance between the adjacent strips satisfies $d=\lambda_0/8$, where $\lambda_0$ is the wavelength in free space, and the distance over the ground plane is set as $h=\lambda_0/6$. The effective radius $r_{\rm eff}$ reads $\lambda_0 /100$. The time dependence is assumed to follow $e^{j \omega t}$. For the main example of finite-sized absorbers, the array size is set to $13.5\lambda_0$, which corresponds to $N = 108$.

\section{Numerical results}
For an infinite array, the required load impedance for perfect absorption, corresponding to various designated incident angles, is calculated by \eqref{eq:periodic_load_imp_den}. 
Considering two specific examples, one for the normal incidence and the other for oblique incidence with $\theta_{\rm i}=80\degree$, the required load impedance $Z_{\rm L}$ for perfect absorption in the designed direction are equal to $ (7.5347 \times 10^4 - j 5.2139\times 10^4)$~$\Omega$/m and  $(1.8921 \times 10^4 - j 1.1154\times 10^5)$~$\Omega$/m, respectively.

\begin{figure}[htbp]
    \centering    \includegraphics{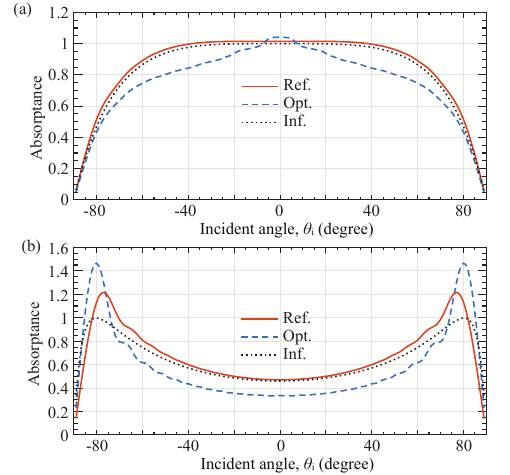}
    \caption{Absorptance as a function of the incident angle $\theta_{\rm i}$ for (a) the design incident angle equals $0^\circ$, and (b) the design incident angle equals $80^\circ$. For finite-sized absorbers, the red solid curve indicates the reference case (all the load impedances are the same as for the designed periodic infinite array), while the blue dashed curve shows the optimized case (the optimized loads). The black dotted curve shows the absorptance of  the designed infinite absorber.}  \label{fig:0_80_abs_ref_opt_inf}
\end{figure}

The absorptance for such an infinite-sized strips array can be calculated analytically as $1-\left|R\right|^2$, since there is no transmission. Here, results are obtained by using commercial software COMSOL Multiphysics. The absorptance as a function of the incident angle is depicted with a black dashed curve in Figs.~\ref{fig:0_80_abs_ref_opt_inf}(a) and (b) for the designed two specific examples. It can be observed that the maximum absorptance is 100\%, which occurs in the designed directions of  $0\degree$ and $80\degree$, respectively. Obviously, the unity absorptance is the maximum attainable value for any infinite passive periodic structure illuminated by a plane wave. 

\begin{figure}[htbp]
    \centering    \includegraphics{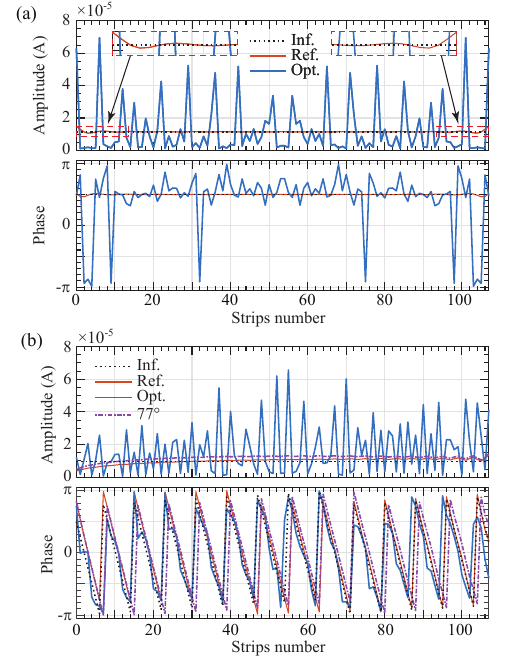}
    \caption{Induced current distribution for two design incident angles with (a) $\theta_{\rm i} = 0\degree$ and (b) $\theta_{\rm i} = 80\degree$. The black dotted curve corresponds to infinite absorbers. For finite-sized absorbers, the red solid curve, and blue solid curve represent the reference case (connected with identical loads), and optimization case (connected with optimized loads), respectively. The purple dot-dashed curve in the subfigure (b) represents the induced currents distribution when the incident plane wave is traveling from $77\degree$.}  
    \label{fig:0_80_current_ref_opt}
\end{figure}

The induced current flowing on the surface of the infinite-sized strips array is calculated by~\eqref{eq:inf_induced_current}. For normal incidence, $I_{\rm inf} = j 1.1494 \times 10^{-5}$A, both the amplitude and phase of the induced current are constant numbers. For oblique incidence at $\theta_{\rm i} = 80\degree$, $I_{\rm inf} = j 9.5579 \times 10^{-6}$A, the amplitude is a constant number while the phase varies  linearly, as  $e^{-j k_0 \sin \theta_{\rm i} y}$. The induced current distribution is depicted in Figs.~\ref{fig:0_80_current_ref_opt}(a) and (b) for $\theta_{\rm i} = 0\degree$ and $80\degree$, respectively. In the perfect absorption regime the specularly reflected wave from the PEC ground plane is eliminated by the field generated by the infinite array of induced currents~\cite{Ra2015thin}, while the power carried by the incident wave is fully dissipated by the lossy load impedances.

In practical designs,  the infinite periodic is trancated into a finite-sized strips array where all elements are loaded by identical impedance loads, calculated from the theory of infinite arrays. The corresponding load impedance is obtained from \eqref{eq:periodic_load_imp_den}. For such a truncated finite-sized strips array, the absorptance defined in \eqref{eq:absorption} as a function of the incident angle $\theta_{\rm i}$ is depicted in~Figs.~\ref{fig:0_80_abs_ref_opt_inf}(a) and (b) for the arrays designed to function as perfect absorbers for the incident angles $\theta_{\rm i} = 0\degree$ and $\theta_{\rm i} = 80\degree$, respectively. We observe that for the design incident angle, the absorptance does not have the unity value. When the incident angle equals $0\degree$ and $80\degree$, the absorptance calculated by \eqref{eq:absorption} for the optimized arrays gives $101.4\%$, and $113.6\%$, respectively. With the increase of the incident angle, the absorptance at the design incident angle shows an increasing trend.

For a better understanding of the super-absorption mechanism, the induced current distributions are depicted in Figs.~\ref{fig:0_80_current_ref_opt}(a) and (b), for $\theta_{\rm i} = 0\degree$ and $80\degree$, respectively. Compared with the infinite array, the induced currents of the finite-sized array are different, especially at the edges of the strips array. 
Let us first compare the performance of infinite and truncated periodic arrays. For  the normal-incidence case, the results on Fig.~\ref{fig:0_80_abs_ref_opt_inf} show that even the reference finite-size absorber exhibits a slightly super-directive (above 100\%) absorptance. This is a counter-intuitive result because in this case, all the array elements are the same as in the corresponding infinite array. This result can be explained by the particularities of the current distribution over the absorber area.  
Over the central area of the array, the amplitude and phase vary slowly and have nearly the same values as in the case of the corresponding infinite periodic array. However, close to the array edges the induced current amplitude is higher, see insets of \figref{fig:0_80_current_ref_opt}(a). Higher current amplitude corresponds to higher absorption in the elements that are close to the edges, resulting in a slight increase of the total absorbed power. Distribution of induced currents in dense wire arrays was studied earlier for semi-infinite arrays of ideally conducting wires in free space (\cite{Rozov} and references therein), where a similar growth of the current amplitude at the edge was also noticed.  However, for normal incidence this effect of increased absorption in regular arrays is rather small and in practice can possibly be neglected.

\begin{figure}[!htbp]
    \centering    \includegraphics{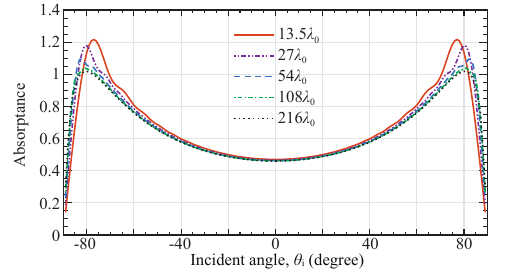}
    \caption{Absorptance as a function of incident angle $\theta_{\rm i}$ for different sized strips array, where the designed angle of incidence is $80\degree$.}
    \label{fig:different_size_abs_ref}
\end{figure}

As is seen from \figref{fig:0_80_current_ref_opt}(b), absorbing arrays designed for oblique angles show much more pronounced super-directive absorption. 
For nearly grazing incidence at $\theta_{\rm i}=80\degree$, the amplitude of the induced currents at a finite-size array shows an increasing trend along the $+y$ direction. The induced currents on the majority of the strips are greater than that observed for the infinite array, particularly toward the ends of the array. The maximum absorptance is 121.6\%, which occurs in the direction of about $\theta_{\rm i}=77\degree$. Note that the load impedance at the corresponding infinite array was found for $\theta_{\rm i}=80\degree$. The induced currents in this case, shown by the purple dot-dashed curve in~\figref{fig:0_80_current_ref_opt}(b), are larger than the currents induced by the wave incident at $\theta_{\rm i}=80\degree$,  which means more dissipated power. The maximum absorptance direction deviation from the design direction is caused by the influence of the element pattern. Although the array factor always aligns accurately with the desired direction, there is a slight shift in the product of the array factor and the element pattern. This shift reduces as the size of the array increases, as the array factor becomes more directive~\cite{li2023all}. This can also be observed and validated in~\figref{fig:different_size_abs_ref} that shows the absorptance as a function of incident angle for absorbers with different sizes. With the increasing length of the strip array, the absorptance of regular arrays decreases. The absorptance will ultimately tend to unity, which is the case of conventional perfect absorption for the infinite structures. This result confirms that the main mechanism of super-absorption in finite-size regular arrays is due to edge effects.

Motivated by the results of \cite{li2023all} that have shown a possibility to realize super-directive anomalous reflectors using subwavelength arrays with optimized loads, next we use optimization techniques to design absorbing arrays. As a reference for comparison, we will use the above discussed results for finite-size absorbers formed by truncation of conventional uniform perfect absorbers.

\begin{figure}[h]
    \centering
    \includegraphics{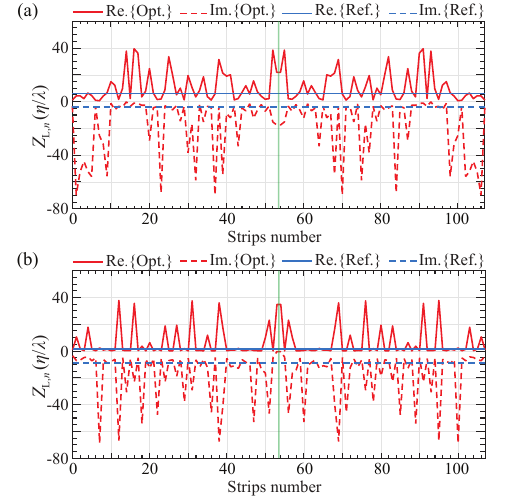}
    \caption{Load impedance distributions for two design incident angles: (a) $\theta_{\rm i} = 0\degree$ and (b) $\theta_{\rm i} = 80\degree$. The red curves show the load impedances of the optimized loads while the blue curves represent the values of the reference load impedance. The solid and dashed curves represent the real and imaginary parts of the load impedance, respectively. The green solid line shows the symmetry axis of the strip array.}   \label{fig:0_80_load_ref_opt}
\end{figure}

Here, the genetic algorithm (GA), a global optimization method, is used to find the optimal load impedances. The objective function is defined as $\mathcal{O} = {\rm min.} \{ -P_{\rm dis} \}$, that is, the goal is to dissipate as much incident power as possible. The absorptance of the optimized arrays as a function of the incident angle is shown by blue dashed curves in~Figs.~\ref{fig:0_80_abs_ref_opt_inf}(a) and (b) for $\theta_{\rm i}=0\degree$ and $80\degree$, respectively. The absorptance for the normal incidence is $104.2\%$, while it reaches as high value as  $146.6\%$ for the large incident angle example. The comparison with the reference case of truncated regular arrays and the optimized array shows that after optimization the absorptance can be significantly improved especially when the incident wave has an extreme incident angle. 
The current distribution in the optimized case becomes highly irregular (see the blue solid curves and black dotted curves in~\figref{fig:0_80_current_ref_opt}), corresponding to high-level of excited evanescent fields in the array vicinity. Also in this case we observe a tendency of higher induced currents close to the array edges. However,  enhanced absorption at the normal incidence leads to weaker absorption at other angles. The load impedances required to realize the required induced currents of the reference case and the optimization case are presented in Figs.~\ref{fig:0_80_load_ref_opt}(a) and (b) for the incident angles of $0\degree$ and $80\degree$, respectively.

\begin{figure}[!htbp]
    \centering
    \includegraphics{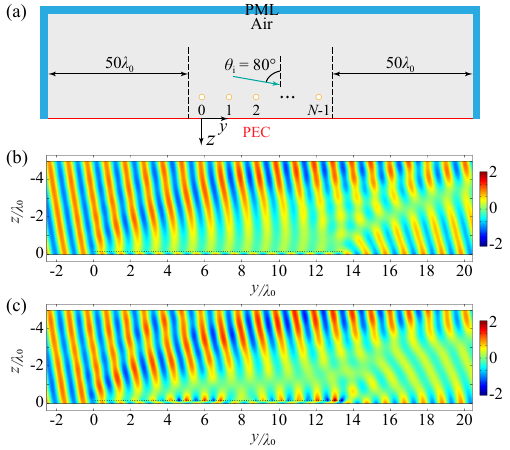}
    \caption{(a) Schematic diagram of the COMSOL Multiphysics simulation. The simulation domain is surrounded by the perfectly matched layer (PML). The array is positioned in the center of the simulation domain. The incident angle of the plane wave equals $80\degree$. Real part of the scattered electric field (with the units [V/m]) distribution when the incident angle equals $80\degree$ for (b) the reference uniform array and (c) for an array with the optimized loads.}
    \label{fig:sca_0_80deg}
\end{figure}

COMSOL Multiphysics is used to calculate the scattered electric field and illustrate the super-absorption effect when the incident angle equals $80\degree$. The configuration of COMSOL Multiphysics is depicted in \figref{fig:sca_0_80deg}(a). The real part of the scattered electric field is depicted in Figs.~\ref{fig:sca_0_80deg}(b) and (c), respectively. Subfigure~\ref{fig:sca_0_80deg}(b) corresponds to the reference case of a truncated uniform strip array (see blue lines in \figref{fig:0_80_load_ref_opt}(b)). Compared with \figref{fig:sca_0_80deg}(b), a wider shadow region can be observed in~\figref{fig:sca_0_80deg}(c), where the strips are loaded by the optimized loads (see red curves in \figref{fig:0_80_load_ref_opt}(b)). A wider shadow means that more power has been dissipated in the lossy loads, that is, absorptance and the effective absorption width are larger. On the surface of the optimized superdirective absorber we see strong surface-wave fields, which confirms that the main mechanism of such superabsorption is the optimized excitation of evanescent waves in the vicinity of the array. Although the geometrical cross-section area is small at oblique angles, the excited surface waves curry power over the whole area of the array where it is dissipated in the optimized resistive loads of the array elements.  

Superdirective arrays usually have narrow frequency bandwidth due to fast variations of the currents and associated high-amplitude reactive fields. Because also in the optimized absorbers we observe fast variations of the induced currents, we next investigate the frequency bandwidth of absorption, comparing the frequency response of the optimized and reference absorbers of different sizes. In order to find the frequency response of the designed structures, one needs to define the frequency dependence of the load impedances as passive bulk loads.
We assume that the load resistances do not depend on the operation frequency, $\Re({Z_{{\rm L}, n}})=R_n$, while the reactive (capacitive) parts of the load impedances  $\Im({Z_{\rm{L}, n}})$ depend on the operation frequency as that of capacitors: $\Im({Z_{\rm {L}, n}})=({-2\pi f C_n})^{-1}$. The load capacitances $C_n$ are found by setting them to the values that correspond to the required reactances $\Im({Z_{\rm {L}, n}})$ at the design frequency $f_0$.  Then, we model the loads in COMSOL Multiphysics as capacitors and resistors, so that the frequency response of the structure can be calculated.

\begin{figure}
    \centering
    \includegraphics{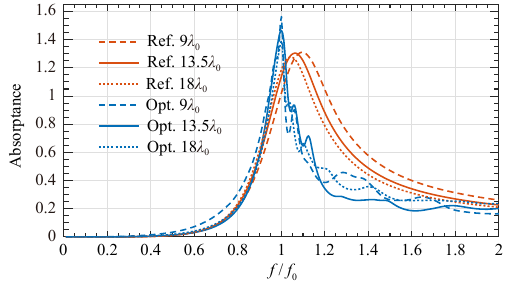}
    \caption{Absorptance as a function of the normalized frequency for arrays of different sizes (indicated in the legend), when the incident angle equals $80\degree$. The results for the reference uniform absorbers are represented by red color, while the curves for the optimized absorbers are represented by blue color.}
    \label{fig:opt_ref_different_size}
\end{figure}

The calculated absorptance as a function of the normalized frequency for different-sized absorbers is depicted in \figref{fig:opt_ref_different_size}. As expected, we see that the optimized superdirective absorbers have a smaller frequency band, but the difference with the reference uniform absorbers is not very large. It is interesting to observe that the frequency of the maximum absorptance of reference structures is shifted from the design frequency.
This is because the effective capacitance of the whole structure decreases when reducing the size of the top reactive layer. The decrease of effective capacitance results in the increase of resonant frequency $\omega_{\rm res}=1/\sqrt{LC}$. 
As the absorber size increases, this frequency shift becomes smaller, as well as the maximum attainable value of absorptance. Ultimately, the response of the reference absorbers tends to the theoretical results for infinite absorbers illuminated by plane waves, with the maximum of absorptance equal to unity occurring at the design frequency $f_0$ that is defined by the effective capacitance and inductance per unit length \cite{Ra2015thin} instead of the parameters of whole finite-size arrays $L,C$. 

\section{Conclusion}
To conclude, we have shown that simple finite-size geometrically periodic arrays can absorb more power than is incident on their surfaces and discussed the physical mechanism of this effect.  Superabsorption is achieved by optimizing the induced surface waves whose strong currents enhance absorption in the array elements. The fields of these waves have fast variations over the array plane. Thus, the key requirement for achieving superdirective absorption is the use of arrays with a subwavelength period, to allow proper control of surface modes. Comparing performance of conventional uniform arrays with the optimized arrays, we have found that especially for arrays designed to absorb waves at near-grazing angles, the absorptance can be significantly increased over the conventional limit of 100\%. For the case of optimized arrays, both the induced currents and load impedances change rapidly over the surface, leading to higher dissipated power. Numerical simulations of scattered fields show a shadow that is wider than the array cross section, as well as the fields of the excited surface waves that enhance the dissipation process over the whole array surface. Although here we considered a simple, analytically solvable example of thin impedance-loaded strips, similar effects can be possibly achieved in arrays of arbitrary small antennas loaded by bulk impedance loads. Importantly, the structure can be reconfigured for superabsorption at any angle by changing the impedances of the loads. We hope that this study not only sheds light on the intriguing phenomenon of super-unity absorptance but also presents a simple possibility for dynamically tuning absorption and enhancing performance of absorbers in real-world applications.


%

\bibliography{manuscript.bib}
\end{document}